\documentclass{article}

\usepackage{arxiv}

\usepackage[utf8]{inputenc} 
\usepackage[T1]{fontenc}    
\usepackage{hyperref}       
\usepackage{url}            
\usepackage{booktabs}       
\usepackage{amsfonts}       
\usepackage{nicefrac}       
\usepackage{microtype}      
\usepackage[title]{appendix}
\usepackage{graphicx}
\usepackage{float}
\usepackage[bf]{caption}

\usepackage{doi}
\usepackage{etoolbox}
\usepackage[natbibapa]{apacite}
\AtBeginDocument{}
\renewcommand{\APACrefnote}[1]{}

\newtoggle{bibdoi}
\newtoggle{biburl}
\makeatletter
\newsavebox{\bib@url}
\newsavebox{\bib@doi}

\undef{\APACrefURL}
\undef{\endAPACrefURL}
\undef{\APACrefDOI}
\undef{\endAPACrefDOI}

\newenvironment{APACrefURL}
  {\global\toggletrue{biburl}\lrbox\bib@url}
  {\endlrbox}

\newenvironment{APACrefDOI}
  {\global\toggletrue{bibdoi}\lrbox\bib@doi}
  {\endlrbox}

\newcommand{\printinfo}{
  \iftoggle{bibdoi}{\usebox{\bib@doi}}{\usebox{\bib@url}}
  \togglefalse{bibdoi}
}

\AtBeginEnvironment{thebibliography}{
\pretocmd{\PrintBackRefs}{%
  \iftoggle{bibdoi}
    {\iftoggle{biburl}{\unskip\unskip}{}\usebox{\bib@doi}}
    {\iftoggle{biburl}{Retrieved from \usebox{\bib@url}}}{}
  \togglefalse{bibdoi}\togglefalse{biburl}%
}{}{}}
\makeatother

\title{Human Mobility Networks Manifest Dissimilar Resilience Characteristics at Macroscopic, Substructure, and Microscopic Scales}

\date{} 					

\renewcommand{\headeright}{}
\renewcommand{\undertitle}{}
\renewcommand{\shorttitle}{}

\hypersetup{
pdftitle={Human Mobility Networks Manifest Dissimilar Resilience Characteristics at Macroscopic, Substructure, and Microscopic Scales},
pdfsubject={},
}

\begin{document}
\maketitle

\begin{center}
{\Large
Chia-Wei Hsu\textsuperscript{a,*},
Matthew Alexander Ho\textsuperscript{a},
Ali Mostafavi\textsuperscript{a}
\par}

\bigskip
\textsuperscript{a} Urban Resilience.AI Lab, Zachry Department of Civil and Environmental Engineering,\\ Texas A\&M University, 199 Spence St., College Station, TX 77843\\
\vspace{6pt}
\textsuperscript{*} correseponding author, email: chawei0207@tamu.edu
\\
\end{center}
\bigskip
\begin{abstract}
Human mobility networks can reveal insights into resilience phenomena, such as population response to, impacts on, and recovery from crises. The majority of human mobility network resilience characterizations, however, focus mainly on macroscopic network properties; little is known about variation in measured resilience characteristics (i.e., the extent of impact and recovery duration) across macroscopic, substructure (motif), and microscopic mobility scales. To address this gap, in this study, we examine the human mobility network in eight parishes in Louisiana (USA) impacted by the 2021 Hurricane Ida. We constructed human mobility networks using location-based data and examined three sets of measures: (1) macroscopic measures, such as network density, giant component size, and modularity; (2) substructure measures, such motif distribution; and (3) microscopic mobility measures, such as the radius of gyration and average travel distance. To determine the extent of impact and duration of recovery, for each measure, we established the baseline values and examined the fluctuation of measures during the perturbation caused by Hurricane Ida. The results reveal the variation of impact extent and recovery duration obtained from different sets of measures at different scales. Macroscopic measures, such as giant components, tend to recover more quickly than substructure and microscopic measures. In fact, microscopic measures tend to recover more slowly than measures in other scales. These findings suggest that resilience characteristics in human mobility networks are scale-variant, and thus, a single measure at a particular scale may not be representative of the perturbation impacts and recovery duration in the network as a whole. These results spotlight  the need to use measures at different scales to properly characterize resilience in human mobility networks.
\end{abstract}



\section{Introduction}
\label{sec:Introduction}

Human mobility networks can reveal insights phenomena such as population response to \citep{fan_evaluating_2021} impacts on \citep{yuan_smart_2022,yuan_unraveling_2022}, and recovery from \citep{lee_specifying_2022} crises, such as natural disasters. Human movements can serve as a medium for understanding the vulnerability and resilience of a community in facing disasters \citep{hsu_human_2022,lee_homophilic_2022,liu_hazard_2022,li_location_2022,rajput_latent_2022}. The expanding availability of high-resolution location-based data has enabled studies linking human mobility networks to their resilience characteristics. The majority of the existing studies focus on macroscopic (global) network measures, such as the giant component size, network density, and network diameter, to specify resilience characteristics to assess the impact of perturbation on human mobility during disasters and pandemics \citep{bonaccorsi_economic_2020,bonaccorsi_evidence_2020,galeazzi_human_2021}.
The sole focus on macroscopic network measures for characterizing human mobility network resilience could be problematic because it provides a limited view of the extent of impact and recovery. Recognizing this, recent studies have recognized the need to examine human mobility networks at all scales (macroscopic, substructure, and microscopic) to provide a more complete characterization of  resilience characteristics \citep{rajput_latent_2022,hsu_limitations_2021}. Macroscopic metrics include
 average node degree, network density, network diameter, giant component size, and network modularity. Substructure metrics are usually related to motif distribution and motif attributes. Microscopic metrics includes average number of trips, average travel distance, average travel time, and average radius of gyration \citep{gonzalez_understanding_2008,hoteit_estimating_2014}. Perturbations caused by natural disasters and other crises could have different manifestations in terms of the impact extent and recovery duration at different scales of human mobility networks. However, the current literature lacks important insights regarding the presence and extent of variation in resilience characteristics of human mobility networks across different scales.
To address this gap, this study focuses on addressing two important questions: (1) The extent to which different network measures within each scale of analysis provide similar indications of the extent of impact and duration of recovery? In other words, within each scale, whether the characterization of resilience properties is sensitive to the measure being used; and (2) The extent to which network measures related to different scales provide consistent indications of the extent of impact and duration of recovery? In other words, whether resilience characteristics vary across different scales. Our hypothesis is that metrics at different levels would react differently to disaster perturbations and yield inconsistent impact extent and recovery durations. If evidence supporting the above hypothesis cannot be rejected, then we may conclude that a single metric cannot capture resilience characteristics in human mobility networks. 
To address the above research questions, we collected data from a high-resolution location-based dataset from Spectus, one of the major location intelligence firms in the United States. Human mobility networks constructed from location-based datasets are commonly used by researchers to capture human movements by performing network analysis at the macroscopic, substructure, and microscopic levels. Our study context is the 2021 Hurricane Ida in the State of Louisiana (USA). In this study, anonymous visit records were transformed into trips at the census-tract level based on their precedence relationships. These trips were aggregated daily to construct human mobility networks and to obtain network measures at different levels. At the macroscopic level, measures examined include average node degree, network density, network diameter, giant component size, and network modularity. At the substructure level, we focus on motif distribution. At the microscopic level, we extract measures of average trip frequency, average travel distance, average travel time and average radius of gyration. For each measure, baseline values are calculated by averaging the value during normal conditions before the hurricane made landfall. Accordingly, we examined the extent of impact and duration of recovery based on fluctuations in the measures caused by the perturbation of the hurricane. Figure 1 illustrates an overview of the study scope and components. 

\begin{figure}
	\centering
    \includegraphics[width=0.9\linewidth]{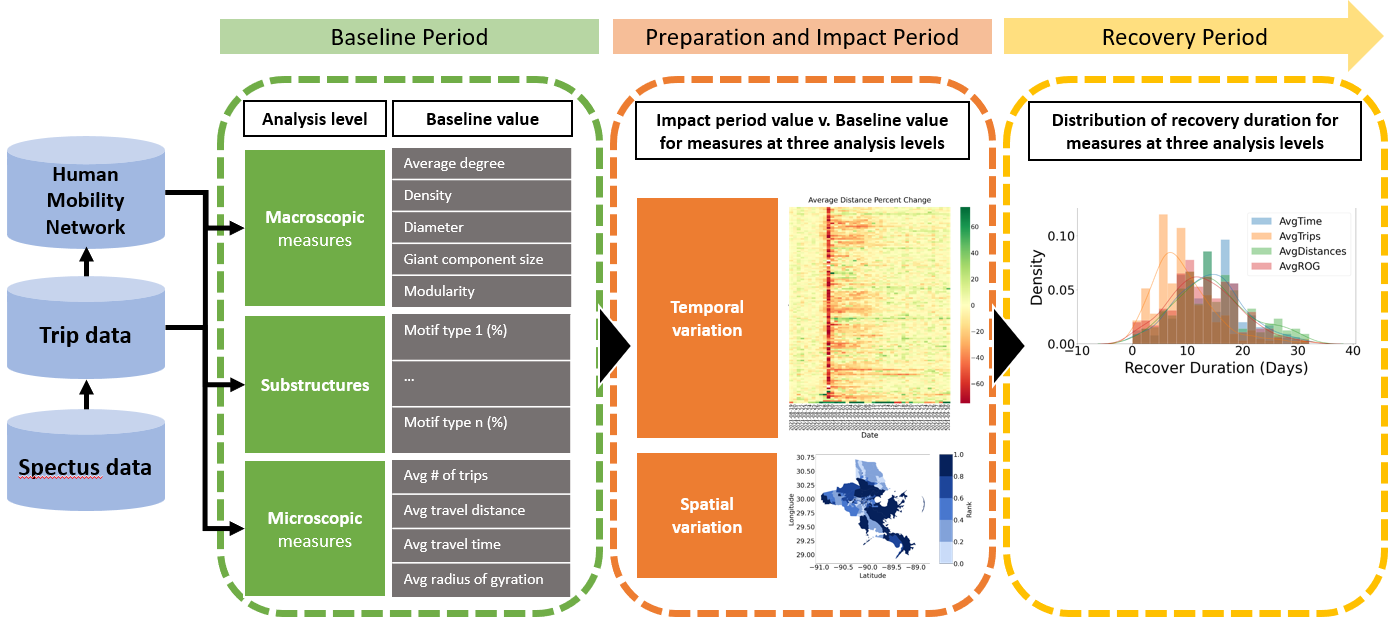}
    \caption{Conceptual illustration of the study scope and components}
	\label{fig:fig1}
\end{figure}

\section{Data description and methods}
\label{sec:Data description and methods}
\subsection{Study context}
To capture the impact and recovery pattern of human mobility networks under the perturbation of disasters, this study examined the change of human mobility network measures at different levels before, during, and after Hurricane Ida in Louisiana. Hurricane Ida was formed on August 26, 2021, and dissipated on September 5, 2021. The period of time before August 26, defined as the baseline period, establishes the level of normal activities without perturbation. The subsequent four days, up until the landfall in Louisiana on August 29, is the preparation period. Residents were informed about the hurricane, and its impact started to increase. The following week, from August 29 to September 5, is the impact period. During this period, impacts of the hurricane became the most intense in Louisiana. The affected area started to recover starting September 5. The anonymized device-level location-based mobility data from August 19, 2021, through August 26, 2021, utilized throughout the network analyses at different levels were collected from Spectus. Human mobility networks in eight parishes in the New Orleans metropolitan area most affected by the hurricane were examined: Jefferson, Orleans, Plaquemines, St. Bernard, St. Charles, St. James, St. John the Baptist, and St. Tammany parishes (Figure 2). (In Louisiana, parishes are the jurisdictional entity corresponding to counties in other states.

\begin{figure}
	\centering
    \includegraphics[width=0.7\linewidth]{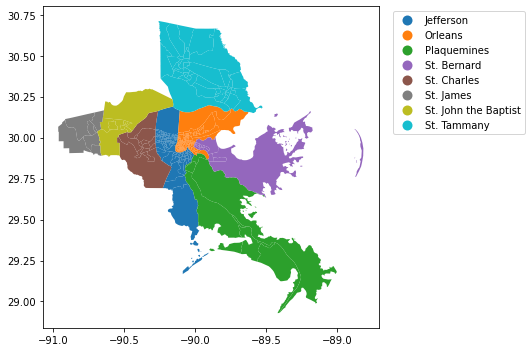}
    \caption{The most affected parishes selected in this study were the parishes of Jefferson, Orleans, Plaquemines, St. Bernard, St. Charles, St. James, St. John the Baptist, and St. Tammany.}
	\label{fig:fig2}
\end{figure}

\subsubsection{Spectus data}
The Global Positioning System (GPS) location dataset used in this study was obtained by Spectus from smartphone devices. Spectus collects large-scale anonymous location data from almost 70 million mobile devices in the United States when users download one of its partner apps and opt in to the app's location services through a General Data Protection Regulation (GDPR)- and California Consumer Privacy Act (CCPA)-compliant framework. Users can choose to opt out of the app's location services at any time. Spectus partners with more than 220 mobile apps that include the proprietary Spectus software development kits (SDK). By collecting data from about one in four US smartphones, Spectus covers almost 20\% of the US population. The data were collected through partner applications and relied on devices’ internal GPS hardware. GPS sensor log data have been previously used as a source of data in fusion frameworks to study human mobility and in travel mode detection since such location data have a high spatiotemporal resolution. In addition to de-identifying data, Spectus applies additional enhancements to preserve privacy, such as obfuscating home locations at the census-block group level and removing sensitive points location of interest (POI) from the data set. The device-level data contain individuals’ recorded location information, including an anonymized individual ID, location, and corresponding time (in seconds). Spectus makes available its own data and processing tools, as well as location-based datasets from other providers, through a data cleanroom environment for privacy preserving analyses of multiple geospatial data assets.

\subsubsection{Baseline calculation}
The baseline period is defined as occurring between August 19 through August 26, 2021. This period is considered as a normal period without any perturbations caused by the hurricane. Therefore, the average values of each human mobility network measure at different levels during this period could serve as a baseline level of normal activities. The difference between measure values during the perturbed period and normal period can indicate the severity of impact. The time needed for measure values to return to a specified level can be viewed as the recovery duration. In our study, we define 100 +/- 10\% as the normal range when measure values recover from above/below the baseline. We define the recovery point as the point at which measure values reach the normal range and no more significant fluctuations occur in the following days. The duration between the end of baseline period (August 26) and this recovery point is defined as the recovery duration.

\subsection{Human mobility network analysis}
The collected data from Spectus were first processed into daily trip counts from origins and destinations of anonymous users. The original form of the data recorded only the anonymized device ID, coordinate, and visit time. For each device, every location of visit is determined based on which census tract polygon it falls in. By the precedence relationship obtained from visit times, a device’s movement, or trajectory is determined. The daily trip counts between census tracts are derived from daily aggregation on a census-tract level. We then use these daily trips counts to construct human mobility networks for analyses at macroscopic and substructure level measures. 
Consider an undirected network represented as $G=(V,E,w)$, where $V$ is the set of nodes, $E$ is the set of edges that connects each of the nodes, and $w$ is the weight assigned to each of the edges. In this study, every node is the centroid of a census tract, and edges will be established if people are traveling from one census tract to another. The variable $w$ is the trip count between origin and destination. A total of 43 daily human mobility networks are generated for each parish. These networks form the basis for macroscopic and substructure-level analyses. 

\subsection{Macroscopic-level analyses}
The macroscopic level mobility measures—average degree, density, diameter, giant component size, and modularity calculated from daily human mobility networks—can all be represented as temporal sequences from August 19 to September 30, 2021, for each census tract in eight selected parishes in Louisiana. 
Average degree is simply the average number of connections per node in the graph, accounting for the strength of connections by incorporating link flows as weights \citep{barabasi_network_2016,newman_scientific_2001}. Nodes with a higher degree or weighted degree may be hubs, such as connecting airports in the transportation network. A higher value of the average degree or weighted degree may also indicate the existence of more significant hubs. Underlying mobility between different regions could be an indicator of the importance of routes and the robustness of the entire network \citep{barrat_architecture_2004,newman_clustering_2001}. Network density, an indicator of connection strength in a network, is computed by calculating the ratio of the number of edges present in the network with the possible number of edges \citep{scott_social_1988}. Network diameter is a measure of the shortest distance between the two most distant nodes in the network. Networks with larger diameters are sparser and have fewer redundant connections, while smaller diameters indicate a more resilient network \citep{zhang_assessing_2015}. Giant component size is an indicator of the number of nodes present in the largest connected component in a network. Modularity measures the degree to which a network’s densely connected nodes can be decoupled into separate communities or modules \citep{newman_modularity_2006}. Increased modularity in a network protects the network against the cascading impacts. In the case of mobility networks, increased modularity would imply that the movements are more localized and happen mostly within different clusters, indicating that movements are restricted; thus increased modularity may not be desirable. 

\subsection{Substructure-level analyses}
Motifs are statistically overrepresented substructures (subgraphs) in a network and have been recognized as simple building blocks of complex networks \citep{artzy-randrup_comment_2004,schneider_unravelling_2013}. A motif $G_m$ is defined as a recurrent multi-node-induced subgraph in network $G$. In this study, we consider motifs constructed with four-node pairs. These four-node pairs are classified into seven motif types (Figure 3) throughout our substructure-level analyses. Motif types 1 and 2 represent densely connected subgraphs where almost all nodes are connected. They represent movement between areas that exhibits the most interconnected movement patterns. Motif type 4 has a three-node cycle and an open edge. Motif types 3 and 5 may represent general commute patterns for work or other lifestyle patterns. Motif type 6 represents a hub-and-spoke structure. Motif type 0 is a symbolic motif type that we use to identify four-node pairs where at least one node is disconnected from its subgraph and does not fall into any of the six types.

\begin{figure}
	\centering
    \includegraphics[width=0.6\linewidth]{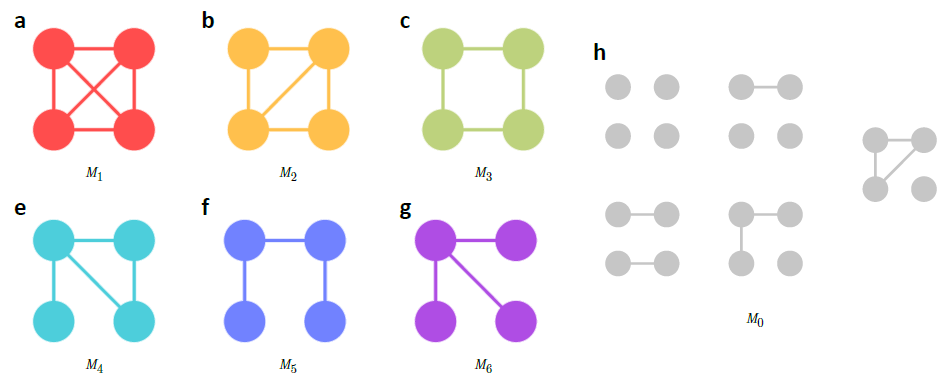}
    \caption{Seven types of four-node pair motifs examined in this study}
	\label{fig:fig3}
\end{figure}

Motif count for each human mobility network was recorded for all parishes from August 19 through September 30, 2021. Previous research working with a different data source noted the difficulty of counting every motif presence because of the size and complexity of networks. They found that sampling 100,000 four-node pairs may accelerate their computation without harming the ability to represent the entire network \citep{rajput_latent_2022}. In this study, we are still able to record every motif presence instead of sampling the four-node pairs. Motif distribution is defined as the number of occurrences of motif types divided by the total number of node pairs. To reduce the noise in the data, we also calculated a seven-day moving average for each day. The key measures we are interested in are the proportion of motif types and their stability. 

\subsection{Microscopic-level analyses}
At the microscopic level, we investigated individual device movements. Each trip of every device is recorded then aggregated to the census-tract level. For each census tract, we filtered the trips that had an origin or destination in that census tract. Then we calculated network characteristics for each census tract based on those filtered trips. The network characteristics we examined at this level were travel volume, travel distance, travel time, and radius of gyration \citep{gonzalez_understanding_2008,hoteit_estimating_2014}. These characteristics are calculated for each census tract daily in each parish.
Travel volume is simply a count of the number of trips originating from a census tract by aggregating trips made by individuals. Higher travel volume indicates a higher activity level. Travel distance is calculated from the averaging the length of trips originating from a census tract. The length of each trip was obtained from the Euclidean distance between centroids of the origin and destination census tracts. This measure is a straightforward indicator that shows whether a census tract is close to a functional center. Travel time is obtained by calculating the difference between the timestamps at the origin and at the destination. This measure usually correlates with travel distance under normal conditions; however, when the road network is disconnected, the same length trip may take longer and thus increase the average travel time. When studying human mobility, the radius of gyration (ROG) is an important statistic that indicates the distance travelled by the device owned by a person during a period. The radius of gyration can be viewed as a proper territory of each device, and thus increasing the territory area means that the person is able to move over longer distances. It is defined as the deviation of device positions from the corresponding centroid position,

given by $r_g=\sqrt{\frac{1}{n}\sum^n_{i=1}(\vec{p_i}-\vec{p_{centroid}})^2}$ , where $\vec{p_i}$ represents the $i$th position recorded for the user and $\vec{p_{centroid}}$ is the center of mass of the user’s recorded displacements obtained as $\vec{p_{centroid}}=\frac{1}{n}\sum^n_{i=1}\vec{p_i}$ \citep{hoteit_estimating_2014,liu_hazard_2022}.

\subsection{Comparing temporal sequences}
By comparing the differences between the temporal sequences of measure values at different levels with their corresponding baselines, we can obtain another set of temporal sequences which represent their percentage of change each day. The sequences can give us insights in terms of the direction and extent of impact by the hurricane and the spatio-temporal pattern of recovery.

\section{Results}
\label{sec:Results}
\subsection{Macroscopic-level analyses}
In this section, we present the results related to the macroscopic measures (i.e., average node degree, density, diameter, giant component size, and modularity). These measures were calculated every day between August 19 and September 30, 2021, for each census tract. Then we compared the resulting temporal sequences with their corresponding baseline values and derived the percentage change of metric values with respect to baseline values. The recovery duration and the distribution of recovery duration were calculated from these temporal sequences based on our definition of recovery point. The range of impact on different metrics are also calculated every day for each parish. 
Figure 4 shows the impact of Hurricane Ida on macroscopic network measures. We can see that the variation can be up to +/-100\% for average degrees while the variations of the other four measures are within +/-40\%. However, there are distinct weekly patterns for each macroscopic network metric under normal conditions. After impacted by Hurricane Ida, the values of all measures flipped along the x-axis and then show recovery by September 30. Figure 5 shows the percentage change of the five measure values across the eight chosen parishes in Louisiana. The results present the sign of changes and extent of impact. Each row of each figure is a macroscopic network measure and each column is a day within our data collection interval. Cell colors with tinges of yellow means neutral or no change, while green and red indicate positive and negative change, respectively. Darker shades of color mean a larger absolute value change. The giant component sizes are almost unchanged, indicating the same set of subparts in the network are still connected. The value of average degree and density decreased while diameter increased during Hurricane Ida. Modularity, on the other hand, was affected but the change is not always positive or negative. This indicates that the network became sparser and possibly more vulnerable against cascading impacts after being affected by Hurricane Ida. Figure 6 shows the ranking by percentile of macroscopic network metrics for the eight parishes chosen in Louisiana showing the impact range across parishes. No region is always assigned the same color. We cannot observe a general trend in metric value change under perturbation; the most affected measures vary across different parishes. 

\begin{figure}
	\centering
    \includegraphics[width=0.8\linewidth]{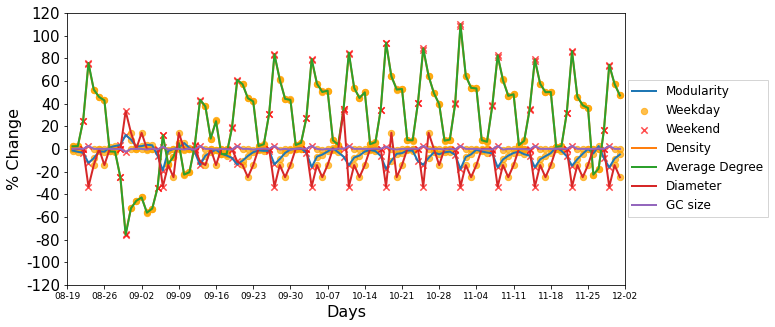}
    \caption{Macroscopic network measures. Impact of Ida on macroscopic network measures such as giant component size, modularity, network density, average node degree, and diameter. Each curve shows the daily fluctuation of measure values. Weekdays and weekends are denoted by crosses and circles, respectively.}
	\label{fig:fig4}
\end{figure}

\begin{figure}
	\centering
    \includegraphics[width=0.8\linewidth]{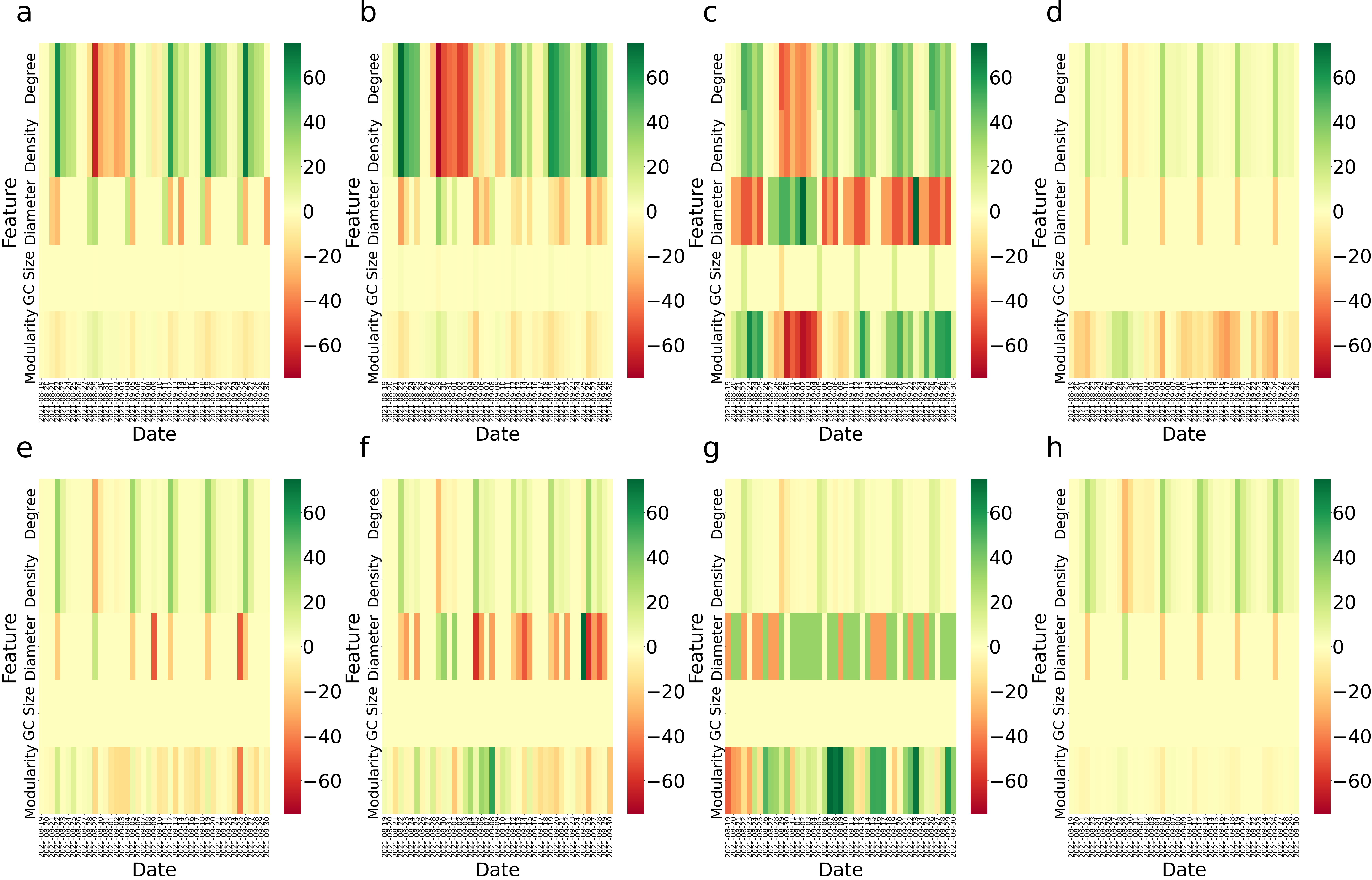}
    \caption{Percentage change of macroscopic metric values with respect to baseline values for the selected eight parishes. (a) Jefferson Parish; (b) Orleans Parish; (c) Plaquemines Parish; (d) St. Bernard Parish; (e) St. Charles Parish; (f) St. James Parish; (g) St. John the Baptist Parish; (h) St. Tammany Parish. Positive change and negative change are represented by red and green shading, while yellow is neutral.}
	\label{fig:fig5}
\end{figure}

\begin{figure}
	\centering
    \includegraphics[width=0.7\linewidth]{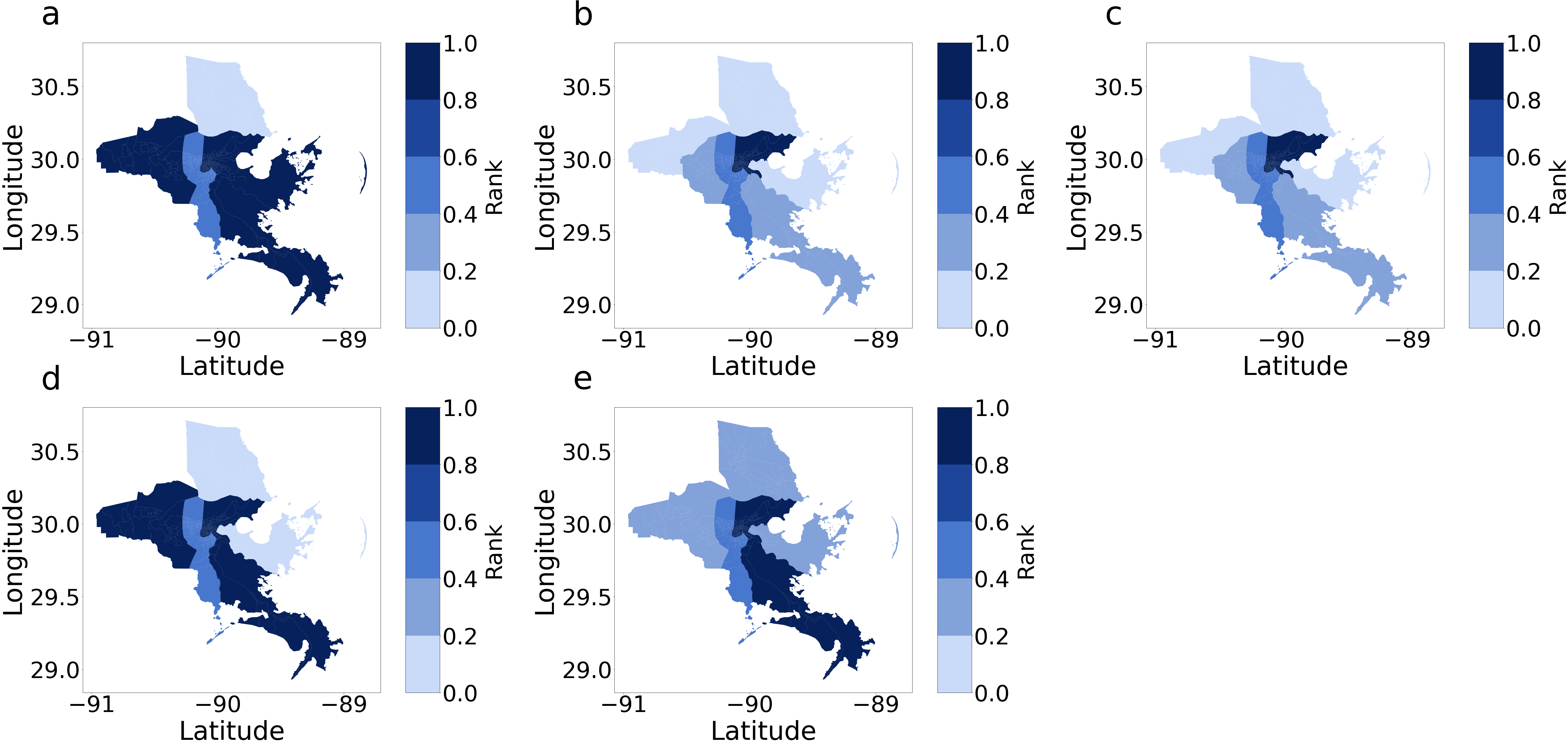}
    \caption{Rank of the impact range of macroscopic network measures for the eight parishes.  (a) average degree; (b) density; (c) diameter; (d) giant component; (e) modularity. This set of figures shows the extent of impact on measures at macroscopic level.}
	\label{fig:fig6}
\end{figure}

\subsection{Substructure level analyses}
In this section, we present the results related to the distribution of motif types and how the proportions change after Hurricane Ida. The proportion of each type of motif is calculated every day between August 19 and September 30, 2012, for each parish. Then we compared the resulting temporal sequences with their corresponding baseline values and derived the percentage change of motif distribution with respect to baseline values. The recovery durations are calculated from these temporal sequences based on our definition of recovery point; then the distribution of recovery duration could be obtained. The range of impact on different motif types is also calculated every day for each parish.
Figure 7 shows the motif distribution for Orleans Parish, Louisiana, during the baseline period, preparation period, impact period, and recovery period. Each sub-figure depicts the motif distribution of a day chosen from each period. Motifs types 4, 5, and 6 account for about 80\% of all motifs. The proportion of major motif types started to fluctuate while the others remained relatively stable during the preparation period. When Hurricane Ida hit the area, only types 4 and 6 remained. In the recovery period, the motif distribution had returned to a similar level as the baseline period. Figure 8 shows the motif distribution change of each motif type for weekdays, weekends, and 7-day moving average for all days. Sometimes the variation of a single day can be up to 250\%, but the curve becomes smoother after we calculated the 7-day moving average. The curves for the major motif types are more stable compared to the other types. Relatively low-volume motifs may be the reason why they look more unstable. It should be noted that not all parishes have the same set of major motif types. The general trend shows a sharp decrease in the number of motifs when affected by Hurricane Ida and then an increase. Another finding is that even when we tracked the motif distribution for a long period after the disaster, most of the motif types still did not completely recover to the baseline level. This may imply that the human mobility network structure at this level takes longer to recover. Figure 9 shows the ranking by percentile of motif distribution in the eight parishes and shows the impact range across parishes. We cannot observe a general trend of motif distribution change under perturbation; the most affected motif types are different across parishes.

\begin{figure}
	\centering
    \includegraphics[width=0.7\linewidth]{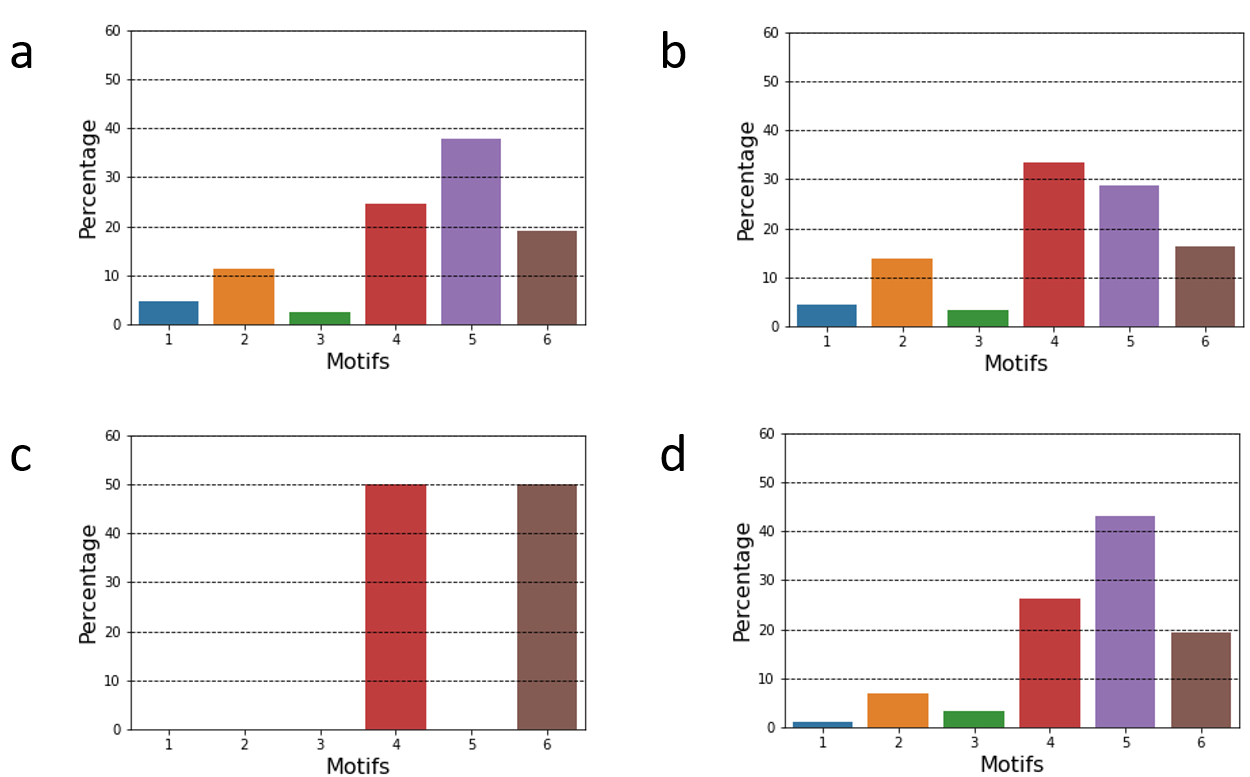}
    \caption{Distribution of motif types in terms of relative occurrence at each stage from daily human mobility networks constructed from datasets for Orleans Parish. (a) baseline period; (b) preparation period; (c) impact period; (d) recovery period. This figure shows the percentage change of motif types under the impact of Hurricane Ida.}
	\label{fig:fig7}
\end{figure}

\begin{figure}
	\centering
    \includegraphics[width=0.8\linewidth]{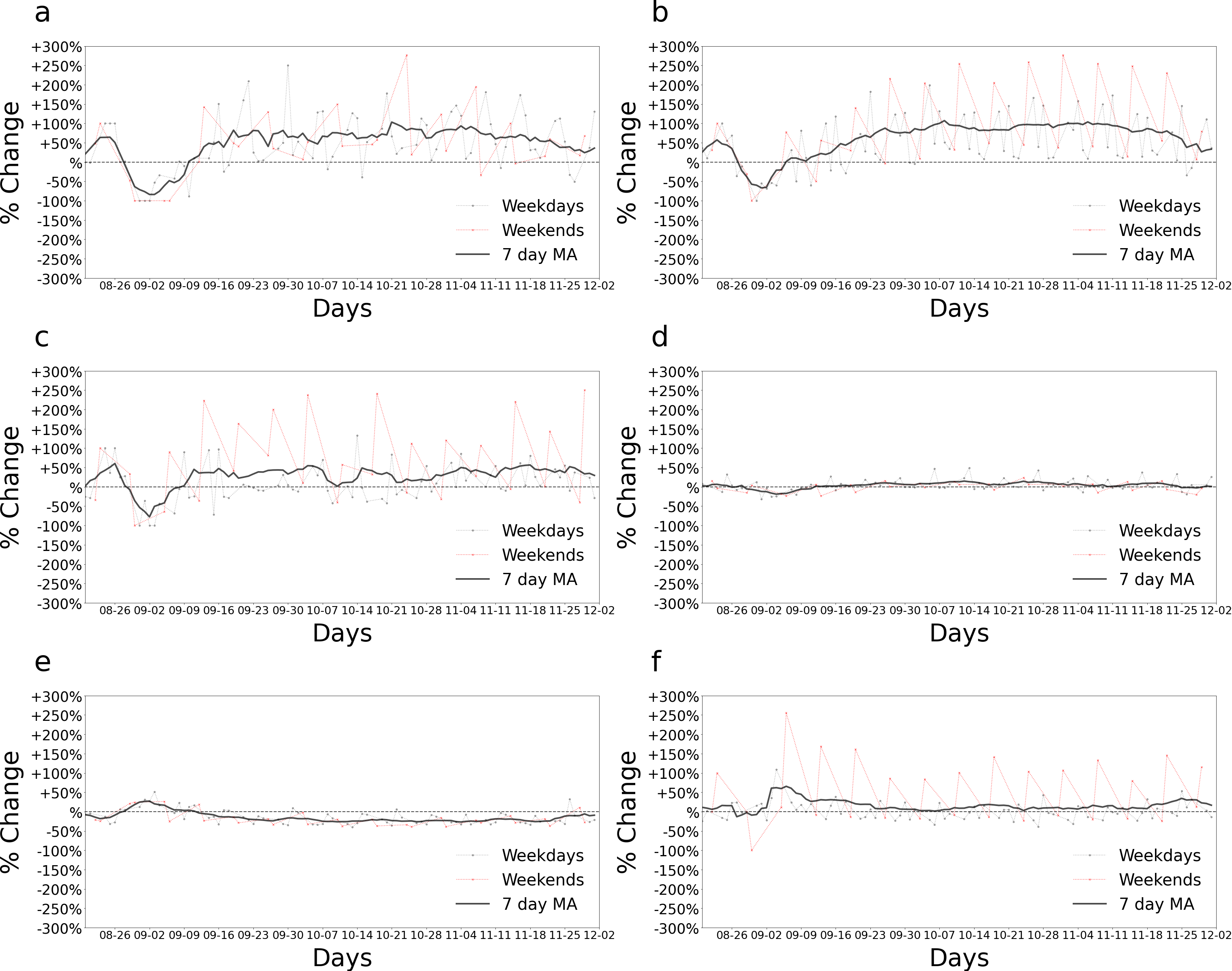}
    \caption{Motif distribution for Orleans Parish. (a)–(f) show the change in motif distribution for motifs 1–6, respectively. Weekday and weekend values are shown in black and grey curves respectively. Black curve is the calculation result of the 7-day moving average without separating weekdays and weekends.}
	\label{fig:fig8}
\end{figure}

\begin{figure}
	\centering
    \includegraphics[width=0.7\linewidth]{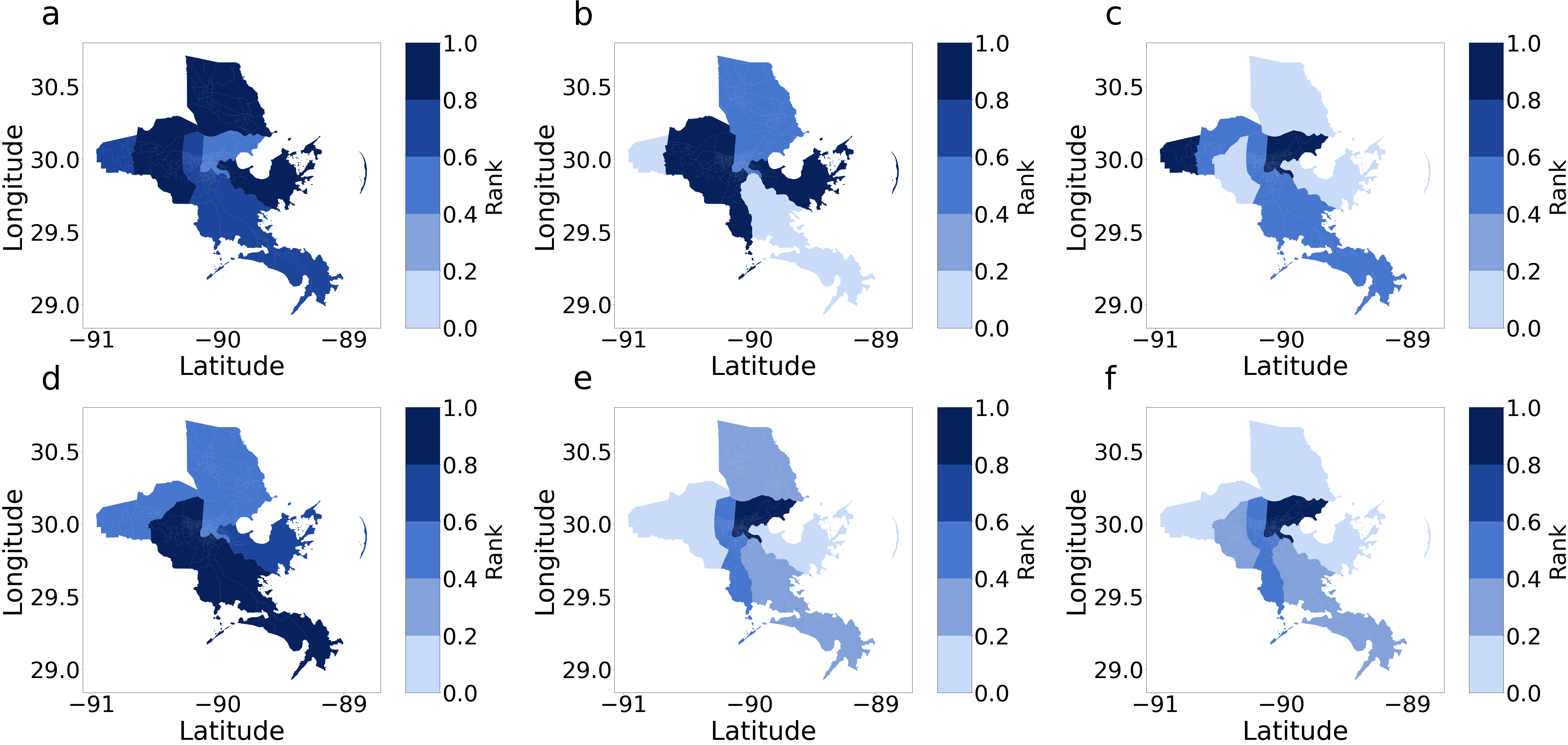}
    \caption{Rank of the impact range of distribution for motif types in the eight parishes. . (a) type 1; (b) type 2; (c) type 3; (d) type 4; (e) type 5; (f) type 6. This set of figures shows the spatial pattern for extent of impact regarding motif distribution at substructural level.}
	\label{fig:fig9}
\end{figure}

\subsection{Microscopic-level analyses}
In this section, we present the results related to the microscopic characteristics of human mobility networks in the face of the hurricane perturbation. Measures for the microscopic characteristics (average number of trips, average trip distance, average travel time, and average radius of gyration) were calculated every day between August 19 and September 30, 2021, for each census tract. Then we compared the resulting temporal fluctuations with their corresponding baseline values and derived the percentage change of metric values with respect to baseline values. The recovery duration and the distribution of recovery duration were calculated from these temporal fluctuations based on our definition of recovery point. The range of impact on different metrics was also calculated every day for each census tract. 
Figure 10 and Figure 11 show the percentage change of the values of the four measures  for Jefferson Parish and Orleans Parish, respectively. They are among the most heavily affected parishes. The figures represent an overview of the evidence of changes and extent of impact. Each row represents  a census tract, and each column, a day within our data collection interval. Cell colors with tinges of yellow means neutral or no change, while green and red indicates positive and negative change, respectively. More saturated color means larger absolute value change. In Figures 7(a) and 7(b) for example, during baseline period, the values of both measures have only slight fluctuations. After Ida struck, the cell color of each census tract turned from red to green in Figure 7(a) indicating that the number of trips first decrease by roughly 40\% then increase by 40\%. In Figure 7(b), the cell color for each census tract turned red, indicating that the travel distances decreased by up to 80\%. Comparing these observations allows us to form an overview about extent and impact and recovery duration across measures. The value of all four measures decreased except for the average number of trips. The average travel time is not the most impacted measure but takes the longest to recover. Average travel distance and average radius of gyration have similar impact and recovery patterns regarding the sign of value change, the extent of impact and the duration of recovery. When we compare the results for Jefferson Parish and Orleans Parish, the recovery may not be complete for some census tracts in Orleans Parish while most of the census tracts in Jefferson Parish had returned to baseline value. Therefore, differences exist in impact percentage and recovery pattern between both measures and parishes. Figure 12 shows the ranking by percentile of census tracts in the eight parishes regarding the impact percentage of measure values. No region is always assigned with the same color except for Orleans Parish. The urban area located at lakeside is most affected when we look at the four microscopic measures. Other than Orleans Parish, we cannot observe a general trend in metric value change under perturbation, the most affected measures vary across parishes.

\begin{figure}[h]
	\centering
    \includegraphics[width=0.8\linewidth]{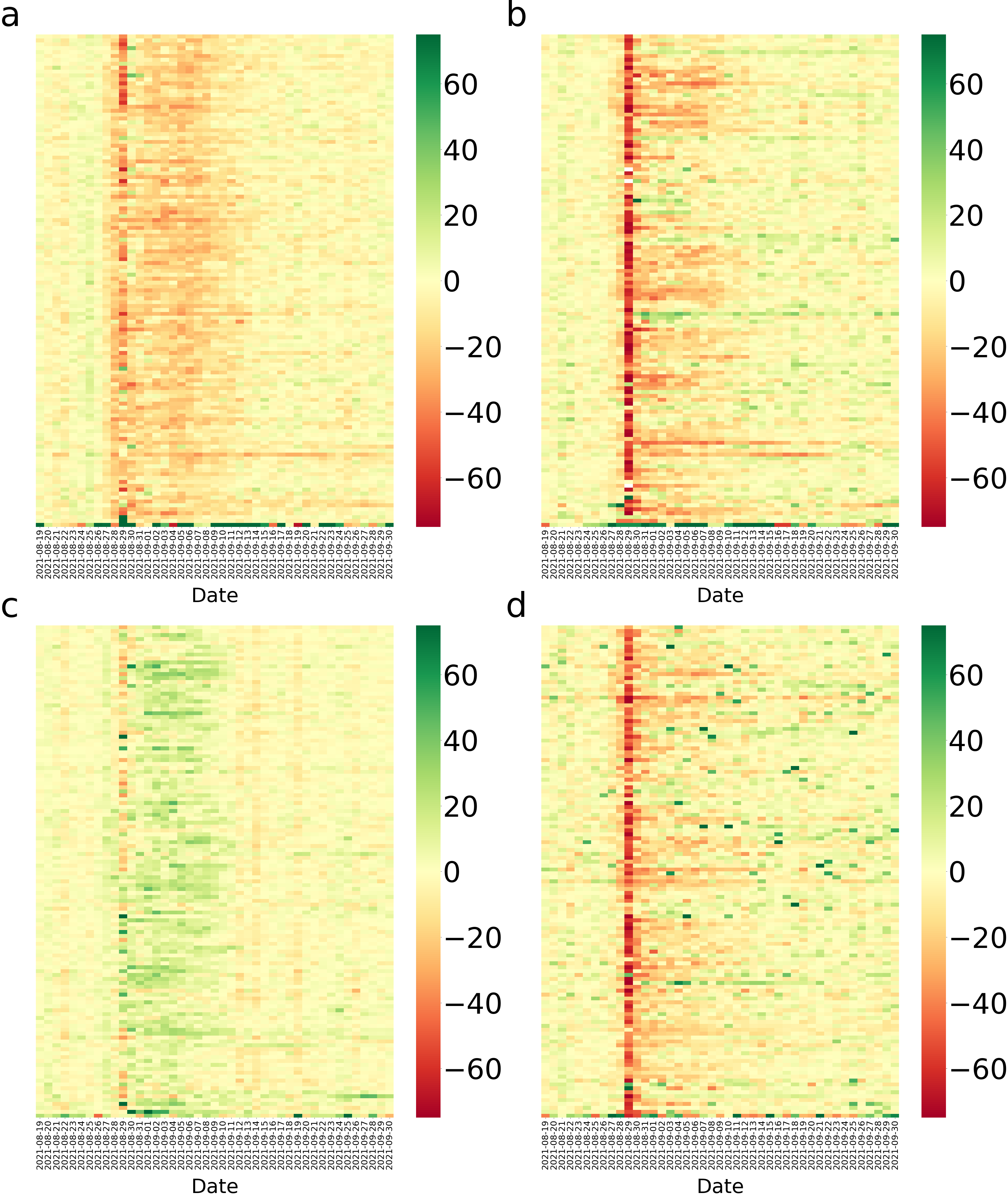}
    \caption{Percentage change of microscopic metric values with respect to baseline values for Jefferson Parish. (a) average number of trips; (b) average travel distance; (c) average radius of gyration; (d) average travel time. Positive change and negative change are represented by red and green shading, while yellow is neutral.}
	\label{fig:fig10}
\end{figure}

\begin{figure}
	\centering
    \includegraphics[width=0.8\linewidth]{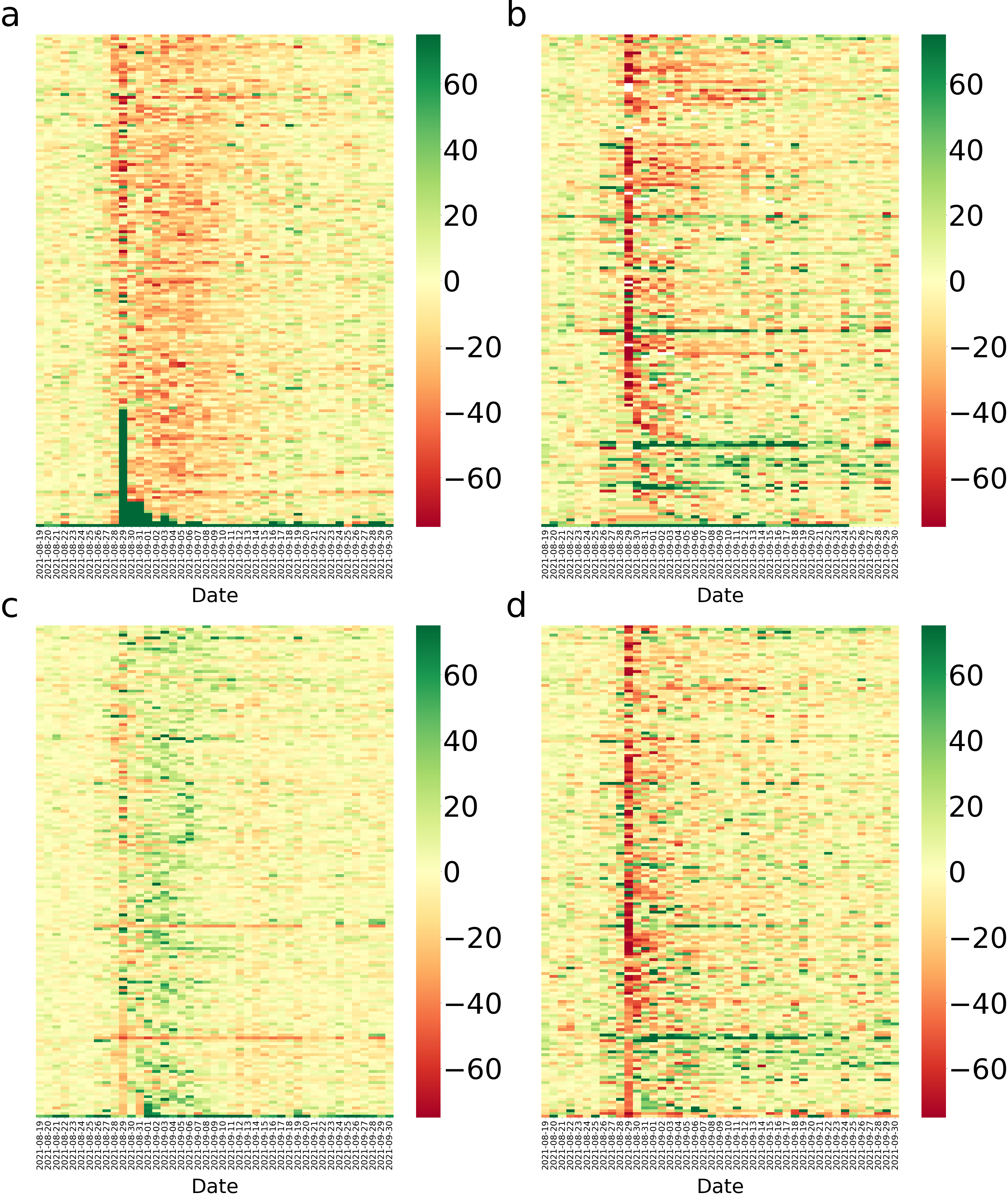}
    \caption{Percentage change of metric values with respect to baseline values for Orleans Parish. (a) average number of trips; (b) average travel distance; (c) average radius of gyration; (d) average travel time. Positive change and negative change are represented by red and green shading, while yellow is neutral.}
	\label{fig:fig11}
\end{figure}

\begin{figure}
	\centering
    \includegraphics[width=0.7\linewidth]{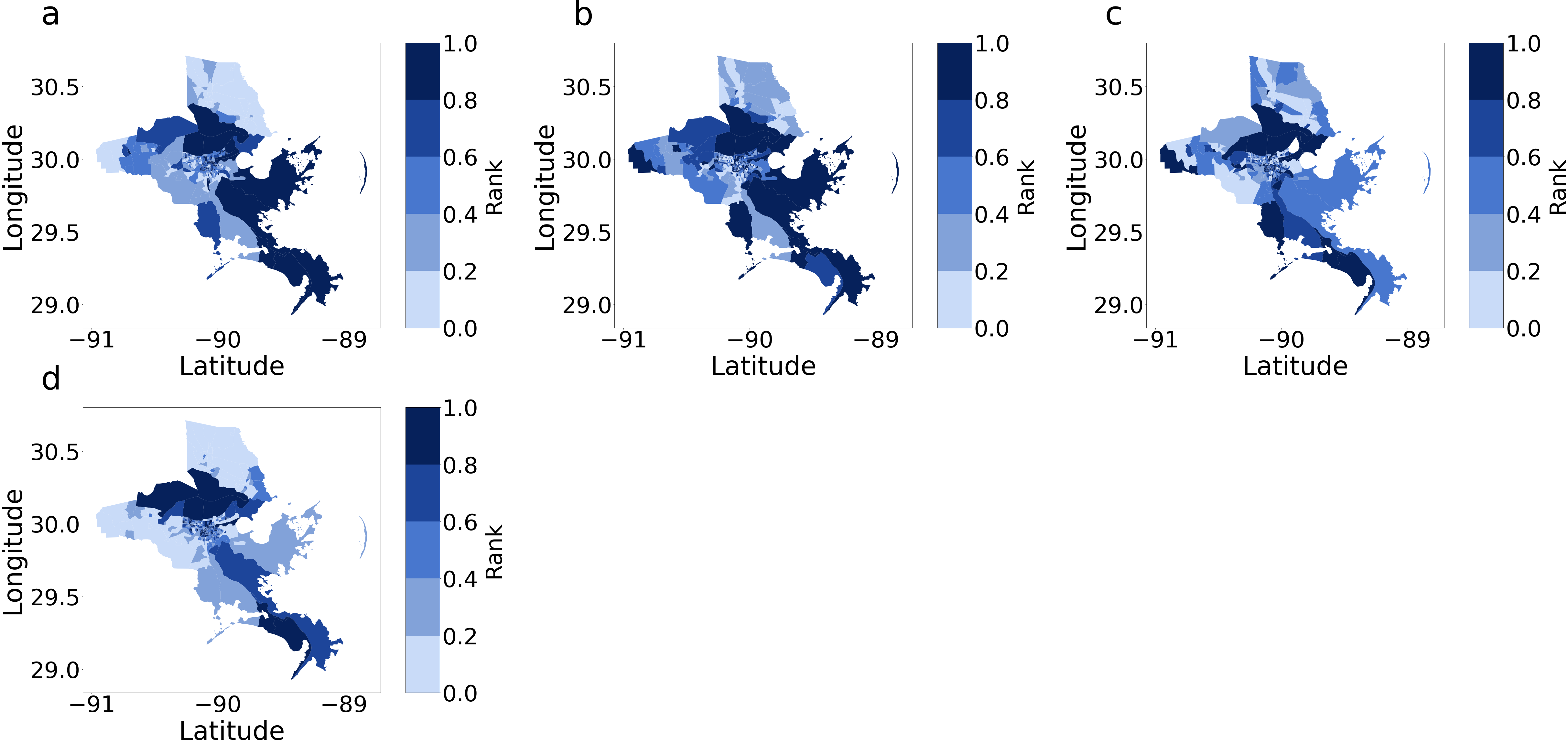}
    \caption{Rank of the impact range of metric values for census tracts in the eight parishes chosen in Louisiana. (a) average number of trips; (b) average travel distance; (c) average radius of gyration; (d) average travel time. This set of figures shows the spatial pattern for extent of impact regarding motif distribution at substructure level.}
	\label{fig:fig12}
\end{figure}

\subsection{Recovery duration across macroscopic, substructure and microscopic-level analyses}
Next, we calculated the recovery duration by our definition at different levels including macroscopic network metrics, motif distribution, and microscopic network measures. The percentage change of each measure was calculated every day; the results are shown in the previous sections. Based on these results, we derived the recovery duration at different levels and obtained their distributions (Figure 13).   This set of graphs was generated from the accumulation of estimated recovery duration of all census tracts in the selected parishes in Louisiana. The histograms were then transformed into density curves to give a better illustration of recovery durations at macroscopic, substructure, and microscopic level. Note that in reality the shortest recovery duration is 0 days. In Figure 12(a), the recovery durations for giant component size are mainly concentrated around 0, which means almost instant recovery. For the other macroscopic measures, the distribution is rather flat, which makes it difficult to conclude a distribution center or mean recovery duration; however, generally the recovery duration of giant component size is longer than the recovery duration of the other macroscopic measures. In Figure 12(b), the distribution of motif 3 is barely changed during Hurricane Ida, giving an instant recovery result. The recovery duration for motif type 6 is also concentrated around 0, which shows shorter recovery time than other motif types. For the rest of the motif types, we may not observe the distinction. In Figure 12(c), average number of trips is right skewed while the others are rather symmetrical. The average number of trips recovers faster than the other three measures. We can summarize that the recovery duration at microscopic level is generally longer than at macroscopic and substructure level. Among macroscopic network metrics, giant component size is the fastest to recover, while the recovery duration distributions for other measures are quite similar to each other. Among motif types, they all recovery quickly without  much difference regarding their recovery duration distribution except for motif type 3 and 6. Among microscopic network metrics, average number of trips is the fastest to recover while, the recovery duration distributions for others are quite similar to each other.

\begin{figure}
	\centering
    \includegraphics[width=0.7\linewidth]{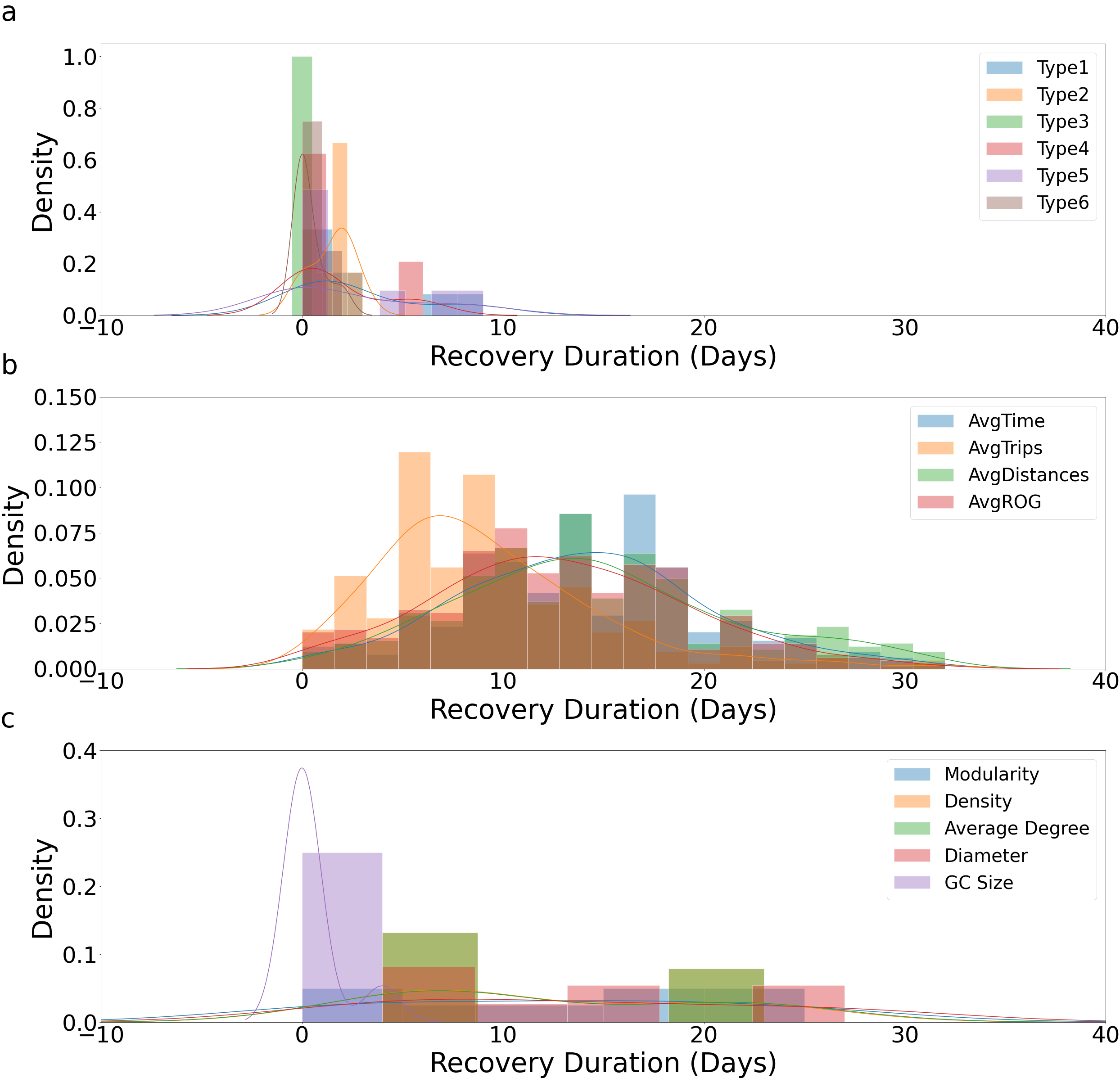}
    \caption{The distribution of recovery duration for network metrics at different levels. (a) macroscopic level; (b) substructure level; (c) microscopic level. This set of figures is generated from the percentage change calculation for each measure at different levels.}
	\label{fig:fig13}
\end{figure}

\section{Discussion and concluding remarks}
\label{sec:Discussion and concluding remarks}
Human mobility networks and their resilience have been the topic of science and engineering studies as an avenue for understanding the dynamics underlying resilience to crises. The majority of existing studies in the literature have primarily focused on macroscopic network measures, such as giant component and network density in characterizing resilience in human mobility networks. 
Location-based datasets processed into human mobility networks for network analyses and metrics at different levels can serve as indicators for critical changes. Not only could human mobility networks created from different datasets yield different results, but metrics at different levels may also have their specific indications. Thus, it is important to understand how network metrics at different levels are affected and to examine their sensitivity under the perturbation of disasters. These analyses in this study help address the question about whether a single metric or analysis result at a single level can be representative when examining the extent and direction of impact and recovery duration. 
The main findings in this study can be summarized in several categories including: (1) the extent and direction of impact of disasters on human mobility network measures at different levels; (2) association between the extent and direction of impact of disasters on human mobility network measures at different scales and spatial areas; and (3) distribution of recovery duration among network measures across different levels. At a macroscopic level, average node degree, density, diameters, giant component size, and modularity are commonly used metrics. In the substructure-level analysis, we examine motif distribution. At the microscopic level, the four chosen metrics are average number of trips, average travel distance, average travel time, and average radius of gyration. The giant component size remains almost unchanged, indicating that the same set of subparts in the network is not affected by the disaster and are still connected. The value of average degree and density decreased while diameter increased during Hurricane Ida. Modularity, on the other hand, was affected, but the direction of impact is not consistent across different parishes. A proportion of major motif types starts to fluctuate while others remain relatively stable during the preparation period. Some motif types cannot be formed during the impact period, but the motif distribution returns to a similar level as the baseline period during the recovery period. The major motif types are not impacted as much as other types. Relatively low volume and difficulty in forming certain types of motifs may be the reason why they seem more unstable. We noted that not all parishes have the same set of major motif types. The general trend shows a sharp decrease in the number of motifs when affected by Hurricane Ida and then an increase as the network recovers. Even after extending the tracking of post-disaster motif distribution, most of the motif types still did not completely recover to the baseline level. This may imply that the human mobility network structure at this level may have experienced more sustained impacts that take longer to recover. The values of all four microscopic network metrics decreased except for the average number of trips. The average travel time is not the most impacted but takes the longest to recover. Average travel distance and average radius of gyration showed similar impact and recovery patterns regarding the direction and extent of impact and the duration of recovery. When we compare the results across different parishes, we found that except for the differences of the impact and recovery pattern between metrics, there is also spatial differences. Regarding  spatial association between the extent and direction of impact of natural disaster on network measures at different levels and parishes, we cannot observe a general trend of metric value change under perturbation at macroscopic and substructure level. The most affected metrics are different across parishes. Certain sets of census tracts tend to receive similar levels of impact at microscopic level. Regarding distribution of recovery duration across network metrics at different levels: the recovery duration at microscopic level is longer than at macroscopic and substructure level. Among macroscopic network metrics, giant component size is the fastest to recover while the recovery duration distribution for others are quite similar. Among motif types, they all recover fast and does not have much difference regarding their recovery duration distribution except for motif type 6. Among microscopic network metrics, average number of trips is the fastest to recover, while the recovery duration distribution for others is quite similar to each other.
These findings highlight the need to understand the impact and recovery of human mobility networks at different levels. The results from this study provide evidence regarding whether it is appropriate to use a single metric to indicate impact level and recovery speed. The significance of the results lies at the necessity of using specific metrics for certain research contexts. Deriving conclusions simply based on a single metric may be somewhat misleading because it can only represent the behavior of only certain types of movements and activities rather than represent a complete picture.  Thus, it is essential to investigate the metrics at different levels which captures a range that covers the possible values regarding the direction and extent of impact as well as recovery duration.

\section*{Acknowledgement}
This material is based in part upon work supported by the National Science Foundation under CRISP 2.0 Type 2 No. 1832662 and the Texas A\&M University X-Grant 699. The authors also would like to acknowledge the data support from Spectus. Any opinions, findings, conclusions, or recommendations expressed in this material are those of the authors and do not necessarily reflect the views of the National Science Foundation, Texas A\&M University, or Spectus.

\section*{Data availability}
All data were collected through a CCPA- and GDPR-compliant framework and utilized for research purposes. The data that support the findings of this study are available from Spectus, but restrictions apply to the availability of these data, which were used under license for the current study. The data can be accessed upon request submitted to the providers. The data was shared under a strict contract through their academic collaborative program, in which they provide access to de-identified and privacy-enhanced mobility data for academic research. All researchers processed and analyzed the data under a non-disclosure agreement and are obligated not to share data further or to attempt to re-identify data.

\section*{Code availability}
The code that supports the findings of this study is available from the corresponding author upon request.

\bibliography{ref}  





\pagebreak

\end{document}